\documentclass[preprint,aps,prd,showpacs,nofootinbib]{revtex4-1}%
\usepackage{epsfig}
\usepackage{amsmath,graphicx}
\usepackage{amsmath}
\usepackage{graphicx}
\usepackage{amsfonts}
\usepackage{amssymb}%
\providecommand{\U}[1]{\protect\rule{.1in}{.1in}}
\begin{document}

\title{Pseudoscalar meson form factors and decays}
\author{A.E. Dorokhov}

\begin{abstract}
In this communication we discuss few topics related with modern experimental
data on the physics of light pseudoscalar mesons. It includes the contribution
of the pseudoscalar mesons to the muon anomalous magnetic moment (AMM), $g-2$, the
rare decays of light pseudoscalar mesons to lepton pair, the transition
form factors of pseudoscalar mesons at large momentum transfer, the pion
transversity form factor.

Measuring the muon anomalous magnetic moment $g-2$ and the rare decays of
light pseudoscalar mesons into lepton pair $P\rightarrow l^{+}l^{-} $ serve as
important test of the standard model. To reduce the theoretical uncertainty in
the standard model predictions the data on the transition form factors of
light pseudoscalar mesons play significant role. Recently new data on behavior
of these form factors at large momentum transfer was supplied by the BABAR collaboration.
Within the nonlocal chiral quark model it shown how to describe these data and how the meson distribution amplitude evolves as a function of the dynamical quark masses and meson masses.

\end{abstract}
\maketitle

\affiliation{Joint Institute for Nuclear Research,
Bogoliubov Laboratory of Theoretical
Physics, \\  141980 Dubna, Moscow region, Russian Federation\\
Institute for Theoretical Problems of Microphysics, Moscow State University, \\  RU-119899, Moscow, Russian Federation}

\section{Introduction}

Among hadrons the pion plays very special role because it has the mass much
lighter than other hadronic states. In theory it appears as a result of
spontaneous breaking of chiral symmetry and gains its small mass due to small
masses of nonstrange current quarks. It seems that confinement forces do not
play any significant role in formation of the pion as a bound state.

Studies of the pion have long history and at the moment the processes with
pion participation is one of standard hadronic background. From other side new
highly-precise experiments related to rare decays or form factors at large
momentum transfer serve as an important independent way to search for physics
beyond the standard model and as test of the QCD dynamics. At the moment,
there are some problems with the matching of the experimental data with the
predictions of the SM. The most famous one is the discrepancy by three
standard deviations between experiment \cite{Bennett:2006fi} and SM theory
\cite{Davier:2010nc,Benayoun:2011mm} for the muon anomalous magnetic moment
(AMM). Another example is similarly large deviation between the recent precise
experimental result on the rare $\pi^{0}$ decay into $e^{-}e^{+}$ pair
\cite{Abouzaid:2006kk} and the SM prediction \cite{Dorokhov:2007bd}.

\section{Muon g-2}

The theoretical studies of the muon AMM $g-2$ (see for review
\cite{Miller:2007kk,Passera:2007fk,Dorokhov:2005ff,Jegerlehner:2009ry,Prades:2009qp}%
), the rare decays of light pseudoscalar mesons into lepton pairs
\cite{Dorokhov:2007bd,Dorokhov:2008cd,Dorokhov:2008qn,Dorokhov:2009xs,Vasko:2011pi}
and the comparison with the experimental results, offer an important
low-energy tests of the SM.

The discrepancy between the present SM prediction of the muon AMM and its
experimental determination \cite{Bennett:2006fi} is $(28.7\pm8.0)\cdot
10^{-10}$ ($3.6\sigma$) \cite{Davier:2010nc}. The theoretical error is
dominated by hadronic corrections. They are the vacuum polarization (HVP) and
light-by-light (LbL) contributions. The latter can be estimated only in model
dependent way. The latest results are the calculations of the pseudoscalar
hadronic channel contribution within the instanton quark model
\cite{Dorokhov:2011zf} and the calculations within the so called
\cite{Goecke:2010if} Dyson-Schwinger equation approach. In
\cite{Dorokhov:2011zf} in the full kinematic dependence of the
meson-two-photon vertices from the virtualities of the mesons and photons is
taken into account. It is demonstrated that the effect of the full kinematic
dependence in the meson-photon vertices is to reduce the contribution of
pseudoscalar exchanges comparing with the most of previous estimates and the
result is $a_{\mu}^{\mathrm{PS,LbL}}=(5.85\pm0.87)\cdot10^{-10}$.

\begin{figure}[h]
\resizebox{0.45\textwidth}{!}{\includegraphics{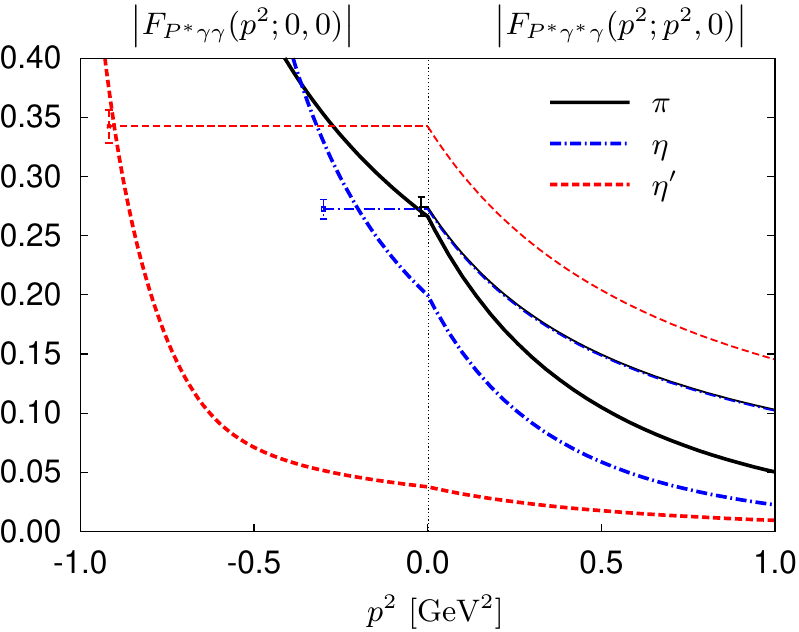}} \caption{Plots of
the $\pi^{0},\eta$ and $\eta^{\prime}$ vertices $F_{P^{\ast}\gamma\gamma
}(p^{2};0,0)$ in the timelike region and $F_{P^{\ast}\gamma^{\ast}\gamma
}(p^{2};p^{2},0)$ in the spacelike region in N$\chi$QM model (thick lines) and
VMD model (thin lines). The points with error bars correspond to the physical
points of the meson decays into two photons. The VMD curves for $\pi^{0}$ and
$\eta$ are almost indistinguishable.}%
\label{fig:CompFF}%
\end{figure}

The knowledge of the eta(') meson couplings to virtual photons is important
for the calculation of the anomalous magnetic moment of the muon, being
pseudoscalar exchange the major contribution to the hadronic light-by-light
scattering. For illustration in Fig. \ref{fig:CompFF} we present on the same
plot the vertex $F_{P^{\ast} \gamma\gamma}(p^{2};0,0)$ in the timelike region
$p^{2}\leq0$ and the vertex $F_{P^{\ast}\gamma^{\ast}\gamma}(p^{2};p^{2},0)$
in the spacelike region $p^{2}\geq0$ as they look in the the instanton quark
model (N$\chi$QM) and vector meson dominance model (VMD). These two special
kinematics match at zero virtuality $p^{2}=0$. The remarkable feature of this
construction is that the first kinematics is connected with the decay of
pseudoscalar mesons into two photons at physical points $F_{P\gamma\gamma
}(-M_{M}^{2};0,0)=g_{P\gamma\gamma}$, while the second kinematics is relevant
for the light-by-light contribution to the muon AMM. Thus, the part of Fig.
\ref{fig:CompFF} at $p^{2}<0$ describes the transition of the pion-two-photon
vertex from the physical points of meson masses to the point with zero
virtuality, which is the edge point of the interval where the integrand is
defined. In VMD type of models there is no such dependence on the meson
virtuality. Thus, the value of this vertex at zero meson virtuality is the
same as the value of the vertex at the physical points of meson masses,
$F_{P\gamma\gamma}^{\mathrm{VMD}} (p^{2}=-M_{\eta,\eta^{\prime}}%
^{2};0,0)=F_{P\gamma\gamma}^{\mathrm{VMD} }(0;0,0)$. However, the $\eta$ and
$\eta^{\prime}$ mesons are much heavier than the pion and such extrapolation
is too crude. One can see that for $\eta$ and particularly for $\eta^{\prime}$
the difference between the values of the vertex at physical and zero
virtuality points is large, $F_{P\gamma\gamma}(p^{2}=-M_{\eta,\eta^{\prime}%
}^{2};0,0)\gg$ $F_{P\gamma\gamma}(0;0,0)$. Thus, the contributions of the
$\eta$ and $\eta^{\prime}$ mesons to the muon AMM evaluated in N$\chi$QM are
strongly suppressed as compared with the VMD results that can only be
considered as upper estimates of these contributions.

In \cite{Goecke:2010if} the hadronic LbL contribution to the muon AMM using
the framework of Dyson-Schwinger equation (DSE) with the result $a_{\mu
}^{\mathrm{PS,qLoop}}=(13.6\pm5.9)\cdot10^{-10}$ that is bigger than most
previous estimates. It is not easy to understand this result, in particular,
in view of discussion in \cite{Boughezal:2011vw}. Moreover, it is instructive
to compare the results on hadronic vacuum polarization contribution in the
instanton \cite{Dorokhov:2004ze} and DSE \cite{Goecke:2011pe} based models. In
\cite{Dorokhov:2004ze} it was demonstrated that the dominant HVP contribution
to muon AMM comes from the dynamical quark loop and contribution of the $\rho$
meson is highly suppressed, while in \cite{Goecke:2011pe} the opposite
conclusion was made. It is still open question what model is more adequate in
these calculations.

\section{Rare pion decay $\pi^{0}\rightarrow e^{+} e^{-}$}

The situation with the rare decays of the light pseudoscalar mesons into
lepton pairs became intriguing after the KTeV E799-II experiment at FermiLab
\cite{Abouzaid:2006kk} in which the pion decay into an electron-positron pair
was measured with high accuracy ($R\left(  P\rightarrow l^{+}l^{-}\right)
=\Gamma\left(  P\rightarrow l^{+}l^{-}\right) /\Gamma_{tot}$)%

\begin{equation}
R^{\mathrm{KTeV}}\left(  \pi^{0}\rightarrow e^{+} e^{-} \right)  =\left(
7.49\pm0.38 \right)  \cdot10^{-8}.\label{KTeV}%
\end{equation}
The standard model prediction gives \cite{Dorokhov:2007bd,Dorokhov:2009xs}
\begin{equation}
R^{\mathrm{Theor}}\left(  \pi^{0}\rightarrow e^{+}e^{-}\right)  =\left(
6.2\pm0.1\right)  \cdot10^{-8},\label{Bth}%
\end{equation}
which is $3.1\sigma$ below the KTeV result (\ref{KTeV}). Other pseudoscalar
mesons rare decays into lepton pair are also well suited as a test of the
standard model (see Table). From experimental point of view the most
interesting are the decays of $\eta$ and $\eta^{\prime}$ to muon pair. The
branchings for these processes theoretically bounded as from below as well
from the above \cite{Dorokhov:2007bd,Dorokhov:2009xs}. Further independent
experiments for $\pi^{0}\to e^{+}e^{-}$ at WASAatCOSY \cite{Kupsc:2008zz} and
for $\eta(\eta^{\prime+}l^{-}$ KLOE \cite{Bloise:2008zz} and BES III
\cite{Li:2009jd} and other facilities will be crucial for resolution of the
problem with the rare leptonic decays of light pseudoscalar mesons. More
precise data on the transition form factors in wider region of momentum
transfer are expected soon from the BABAR, BELLE (at large momentum transfer)
and KEDR, KLOE \cite{AmelinoCamelia:2010me} (at small momentum transfer)
collaborations. These data would allow to make more accurate theoretical predictions.

\begin{table}[th]
\caption[Results]{Values of the branchings $R\left(  P\rightarrow l^{+}%
l^{-}\right) =\Gamma\left(  P\rightarrow l^{+}l^{-}\right) /\Gamma_{tot}$
obtained in approach \cite{Dorokhov:2007bd,Dorokhov:2009xs} and compared with
the available experimental results. }%
\label{table2}
\renewcommand{\arraystretch}{1.0}
\begin{tabular}
[c]{|c|c|c|c|c|c|}\hline
$R$ & Unitary & CLEO+BABAR & CLEO+BABAR & With mass & Experiment\\
& bound & bound & +OPE & corrections & \\\hline
$R\left(  \pi^{0}\rightarrow e^{+}e^{-}\right)  \times10^{8}$ & $\geq4.69$ &
$\geq5.85\pm0.03$ & $6.23\pm0.12$ & $6.26$ & $7.49\pm0.38$
\cite{Abouzaid:2006kk}\\\hline
$R\left(  \eta\rightarrow\mu^{+}\mu^{-}\right)  \times10^{6}$ & $\geq4.36$ &
$\leq6.60\pm0.12$ & $5.35\pm0.27$ & $4.76$ & $5.8\pm0.8$
\cite{Nakamura:2010zzi,Abegg:1994wx}\\\hline
$R\left(  \eta\rightarrow e^{+}e^{-}\right)  \times10^{9}$ & $\geq1.78$ &
$\geq4.27\pm0.02$ & $4.53\pm0.09$ & $5.19$ & $\leq2.7\cdot10^{4}$
\cite{Berlowski:2008zz}\\\hline
$R\left(  \eta^{\prime}\rightarrow\mu^{+}\mu^{-}\right)  \times10^{7}$ &
$\geq1.35$ & $\leq1.44\pm0.01$ & $1.364\pm0.010$ & $1.24$ & \\\hline
$R\left(  \eta^{\prime}\rightarrow e^{+}e^{-}\right)  \times10^{10}$ &
$\geq0.36$ & $\geq1.121\pm0.004$ & $1.182\pm0.014$ & $1.83$ & \\\hline
\end{tabular}
\end{table}

There are quite few attempts in the literature, to explain the excess of the
experimental data on the $\pi^{0}\rightarrow e^{+}e^{-}$ decay over the SM
prediction, as a manifestation of physics beyond the SM. In Ref.
\cite{Kahn:2007ru}, it was shown that this excess could be explained within
the currently popular model of light dark matter involving a low mass
($\sim10$ MeV) vector bosons $U_{\mu}$, which presumably couple to the
axial-vector currents of quarks and leptons. Another possibility was proposed
in Ref. \cite{Chang:2008np,McKeen:2008gd}, interpreting the same experimental
effect as the contribution of the light CP-odd Higgs boson appearing in the
next-to-minimal supersymmetric SM. The latter version perhaps excluded by
modern experiments.

\section{Pseudoscalar meson transition form factors}

The main limitation on realistic predictions for above processes originates
from the large distance contributions of the strong sector of the SM, where
perturbative QCD does not work. In order to diminish the theoretical
uncertainties, the use of the experimental data on the pion charge and
transition form factors are of crucial importance. The first one, measured in
$e^{+}e^{-} \to\pi^{+}\pi^{-}(\gamma)$ by CMD-2 \cite{Akhmetshin:2006bx}, SND
\cite{Achasov:2006vp}, KLOE \cite{Aloisio:2004bu}, and BABAR \cite{:2009fg}
provides an estimate for the hadron vacuum polarization contribution to muon
$g-2$, with accuracy better than $1\%$. The second one, measured in
$e^{+}e^{-} \to e^{+}e^{-}P$ for spacelike photons by CELLO
\cite{Behrend:1990sr}, CLEO \cite{Gronberg:1997fj}, and BABAR
\cite{:2009mc,:2011hk,Lees:2010de} collaborations and in $e^{+}e^{-} \to
P\gamma$ for timelike photons by the BABAR \cite{Aubert:2006cy} collaboration,
is essential to reduce the theoretical uncertainties in the estimates of the
contributions of the hadronic light-by-light process to the muon $g-2$ and in
the estimates of the decay widths of $P\to l^{+}l^{-}$. The BABAR data
\cite{Aubert:2006cy,:2009mc} on the large momentum behavior of the form
factors cause the following problems for their theoretical interpretation
\cite{Dorokhov:2009jd,Dorokhov:2009zx}: 1) An unexpectedly slow decrease of
the pion transition form factor at high momenta \cite{:2009mc}, 2) the
qualitative difference in the behavior of the pion and $\eta,\eta^{\prime}$
form factors at high momenta \cite{:2011hk}, 3) inconsistency of the measured
ratio of the $\eta,\eta^{\prime}$ form factors with the predicted one
\cite{Aubert:2006cy}.

\begin{figure}[th]
\hspace*{-10mm}\includegraphics[width=0.6\textwidth]{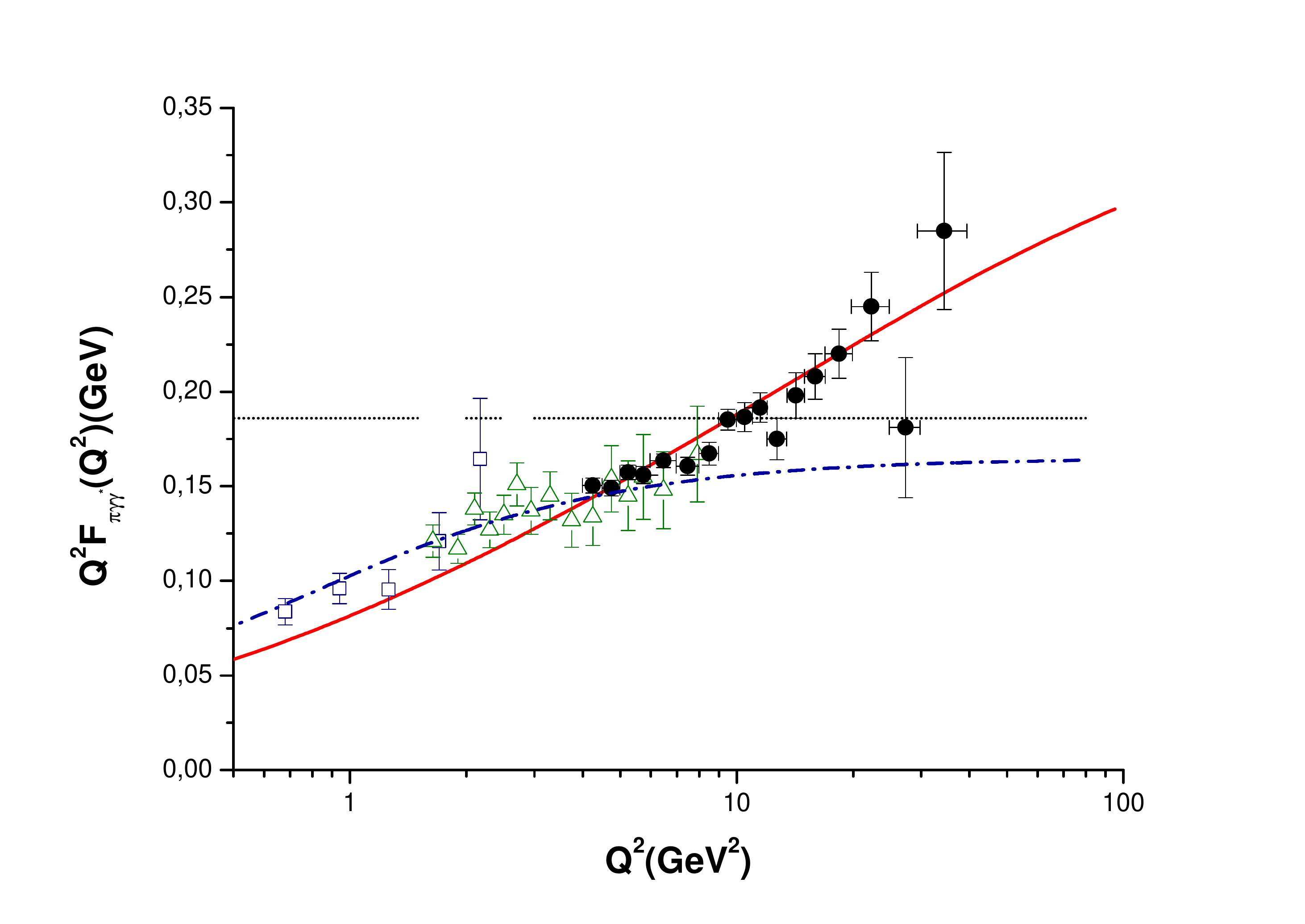} \vspace
*{-10mm}\caption{{\protect\footnotesize The transition form factor
$\gamma^{\ast}\gamma\rightarrow\pi^{0}$. The data are from the CELLO
\cite{Behrend:1990sr}, CLEO \cite{Gronberg:1997fj} and BABAR \cite{:2009mc}
Collaborations. The solid line is the nonlocal chiral quark model calculations and the dash-dot line is the parametrization (6). The dotted line is massless QCD asymptotic limit.}}%
\label{fig:pi}%
\end{figure}

\begin{figure}[th]
\hspace*{-10mm}\includegraphics[width=0.6\textwidth]{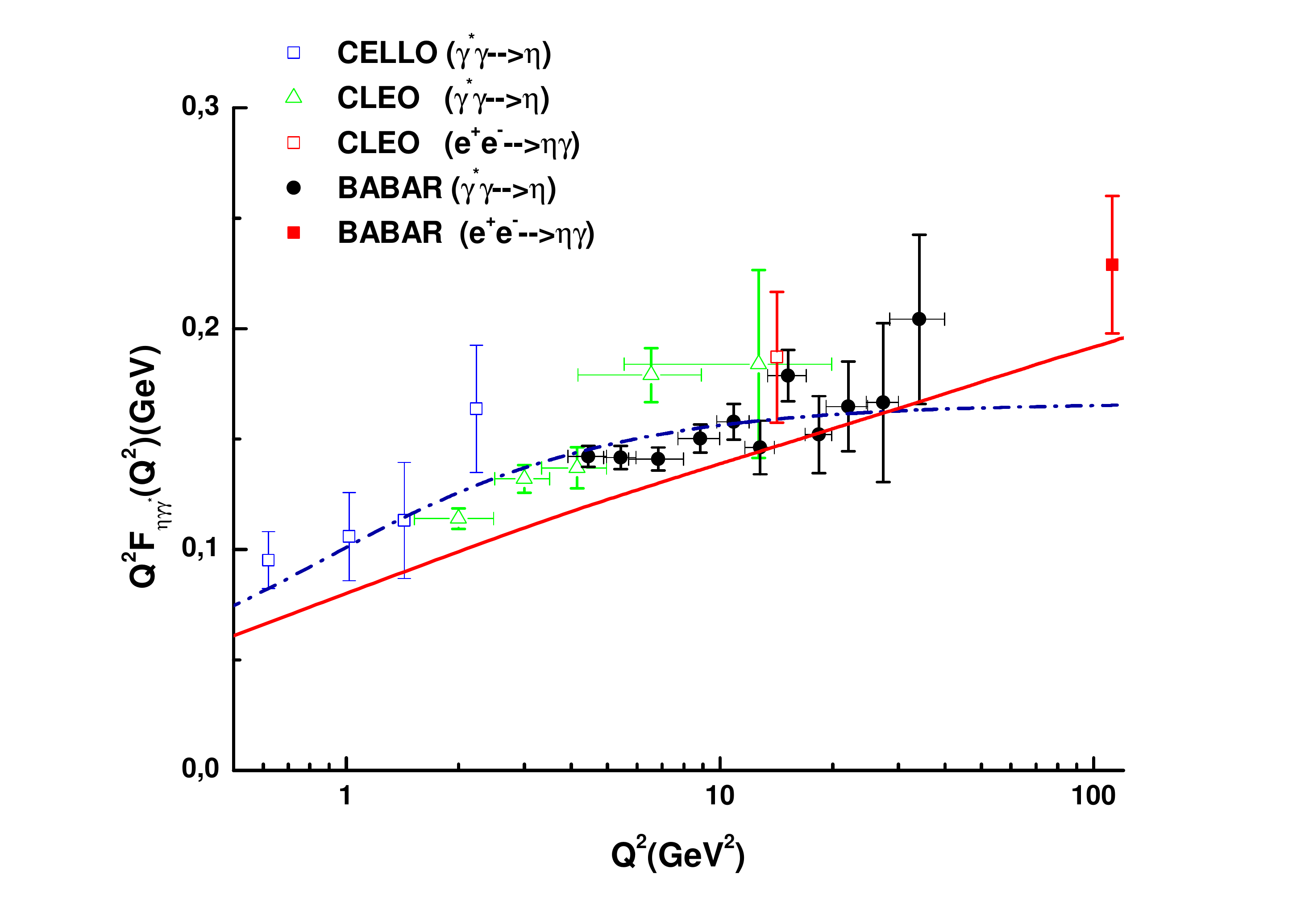} \vspace
*{-10mm}\caption{{\protect\footnotesize The transition form factor
$\gamma^{\ast}\gamma\rightarrow\eta$. The data are from the CELLO
\cite{Behrend:1990sr}, CLEO \cite{Gronberg:1997fj} and BABAR \cite{:2011hk}
Collaborations. The CLEO results obtained in different $\eta$ decay modes are
averaged. The CLEO and BABAR points, measured in the timelike region
$\gamma^{\ast}\rightarrow\eta\gamma$ \cite{:2009tia} and \cite{Aubert:2006cy},
are drawn at $Q^{2}=14.2$ GeV$^{2}$ and $Q^{2}=112$ GeV$^{2}$,
correspondingly, assuming that the spacelike and timelike asymptotics of the
form factor are similar. The solid line is the nonlocal chiral quark model calculations and the dash-dot line is the parametrization (6).  }}%
\label{fig:eta}%
\end{figure}

\begin{figure}[th]
\hspace*{-10mm}\includegraphics[width=0.6\textwidth]{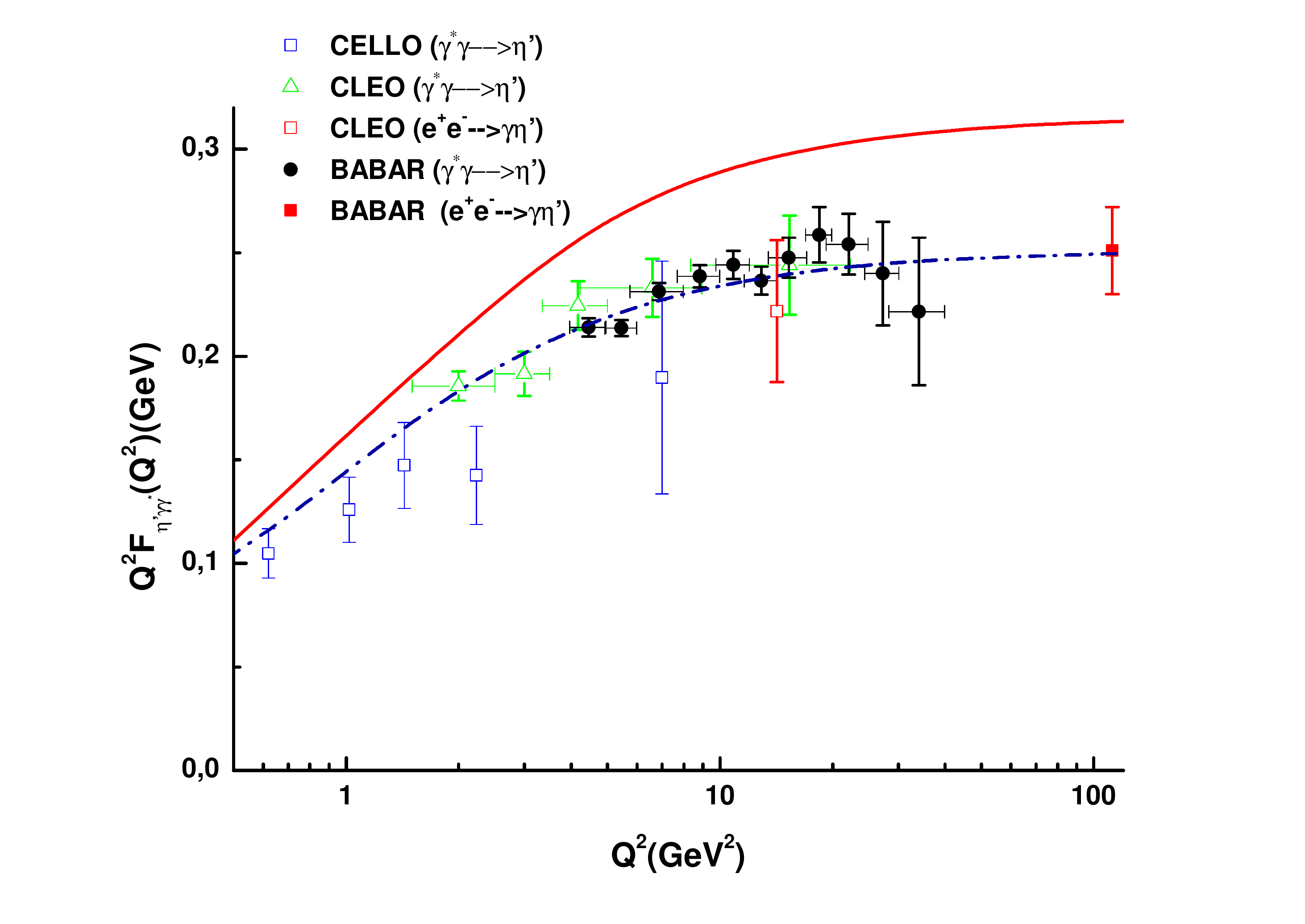}
\vspace*{-10mm}\caption{{\protect\footnotesize The transition form factor
$\gamma^{\ast}\gamma\rightarrow\eta^{\prime}$. The data are from the CELLO
\cite{Behrend:1990sr}, CLEO \cite{Gronberg:1997fj} and BABAR \cite{:2011hk}
Collaborations. The CLEO results obtained in different $\eta^{\prime}$ decay
modes are averaged. The CLEO and BABAR points, measured in the timelike region
$\gamma^{\ast}\rightarrow\eta\gamma$ \cite{:2009tia} and \cite{Aubert:2006cy},
are drawn at $Q^{2}=14.2$ GeV$^{2}$ and $Q^{2}=112$ GeV$^{2}$,
correspondingly, assuming that the spacelike and timelike asymptotics of the
form factor are similar. The solid line is the nonlocal chiral quark model calculations and the dash-dot line is the parametrization (6). }}%
\label{fig:eta1}%
\end{figure}

\begin{figure}[th]
\hspace*{-10mm}\includegraphics[width=0.6\textwidth]{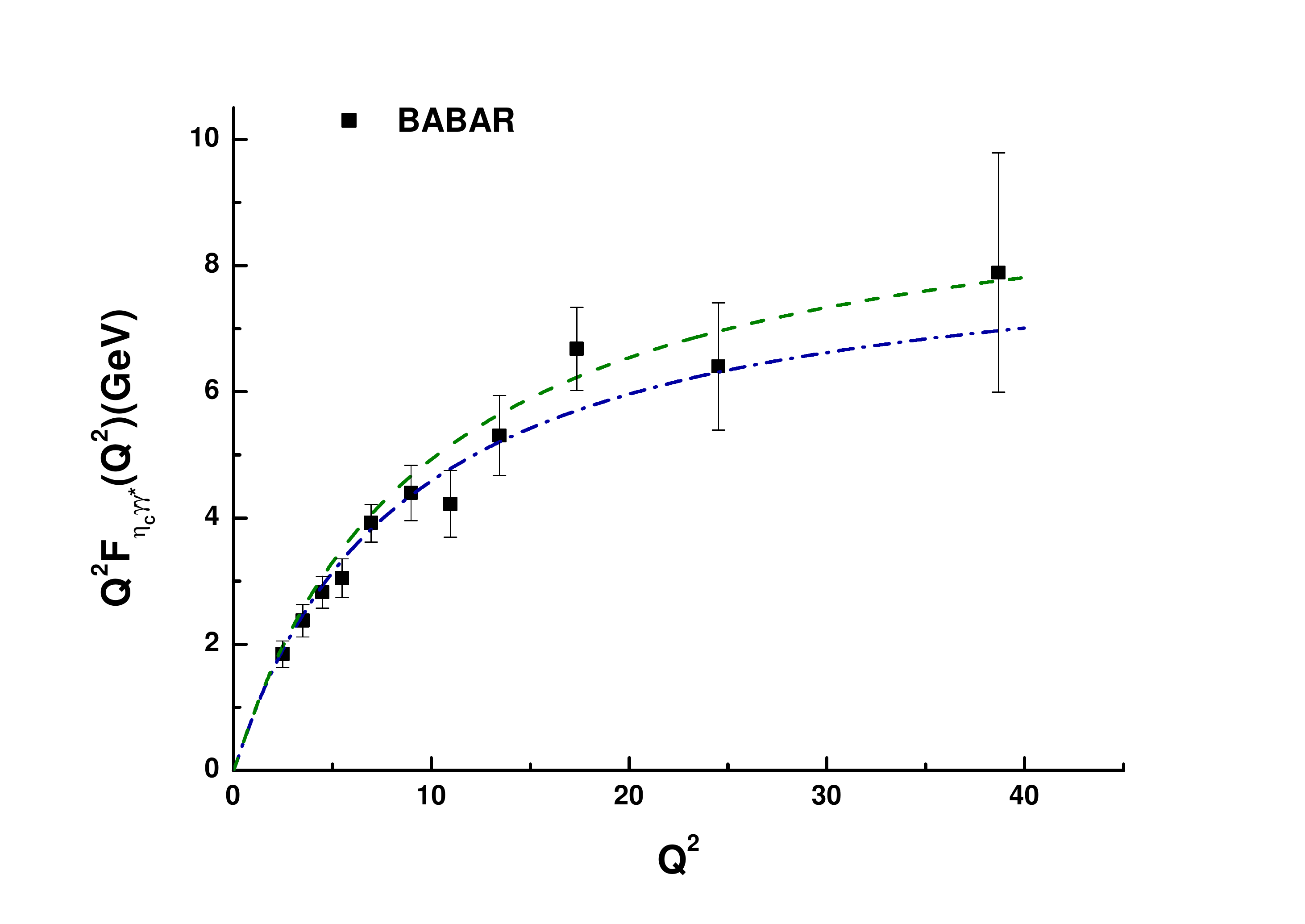}
\vspace*{-10mm}\caption{{\protect\footnotesize The transition form factor
$\gamma^{\ast}\gamma\rightarrow\eta_{C}$. The data are from the BABAR
\cite{Lees:2010de} Collaboration. The dashed line is the nonlocal model
calculation and the dot-dashed line is the fit. (The normalization is
arbitrary)}}%
\label{fig:etaC}%
\end{figure}

In Figs. \ref{fig:pi}, \ref{fig:eta}, \ref{fig:eta1} and \ref{fig:etaC} the
data for the $\pi^{0}$, $\eta$, $\eta^{\prime}$ and $\eta_{c}$ transition form
factors from the CELLO, CLEO, and BABAR collaborations are presented. In Figs.
\ref{fig:eta} and \ref{fig:eta1}, the CLEO \cite{:2009tia} and BABAR
\cite{Aubert:2006cy} points, measured in the timelike region, are drawn at
$Q^{2}=14.2$ GeV$^{2}$ and $Q^{2}=112$ GeV$^{2}$, correspondingly, assuming
that the spacelike and timelike asymptotics of the form factor are equal. It
is seen from the Figs. \ref{fig:eta} and \ref{fig:eta1}, that the spacelike
and timelike points are well conjugated.


At zero momentum transfer, the transition form factor is fixed by the
two-photon decay width (see for recent discussion \cite{Belicka:2011ui})
\begin{equation}
F_{P\gamma\gamma}^{2}(0,0)=\frac{1}{(4\pi\alpha)^{2}}\frac{64\pi\Gamma
(P\to\gamma\gamma)}{M_{P}^{3}},\label{F0exp}%
\end{equation}
where $\alpha$ is the QED coupling constant, $M_{P}$ is the resonance mass and
$\Gamma(P\to\gamma\gamma)$ is the two-photon partial width of the meson $P$.
The axial anomaly predicts
\begin{equation}
F_{P\gamma\gamma^{*}}(Q^{2}=0,0)\approx\frac{1}{4\pi^{2}f_{P}},\label{F0the}%
\end{equation}
where $f_{P}$ is the meson decay constant. Under assumption of factorization,
perturbative QCD predicts the asymptotic behavior of the $F_{P\gamma\gamma
^{*}}^{2}(Q^{2},0)$ transition form factors as $Q^{2}\to\infty$
\cite{Brodsky:1981rp}
\begin{equation}
F_{P\gamma\gamma^{*}}(Q^{2}\to\infty,0)\sim\frac{2f_{P}}{Q^{2}}.\label{Fas}%
\end{equation}
The perturbative QCD corrections to this expression at large momentum transfer
are quite small
\cite{Chase:1979ck,Braaten:1982yp,Kadantseva:1985kb,Mikhailov:2009kf}. The
CLEO (and CELLO) collaboration parameterized their data by a formula similar
to that proposed by Brodsky and Lepage in \cite{Brodsky:1981rp}, but with the
pole mass being a free fitting parameter \cite{Gronberg:1997fj},
\begin{equation}
F^{\mathrm{{CLEO}}}_{\pi\gamma\gamma^{*}}(Q^{2},0)=\frac{1}{4\pi^{2}f_{P}%
}\frac{1}{1+Q^{2}/\Lambda_{P}^{2}},\label{Fcleo}%
\end{equation}
where the values of $f_{P}$ are estimated from (\ref{F0exp}) and (\ref{F0the})
\cite{Gronberg:1997fj}: $f_{\pi}=92.3$ MeV, $f_{\eta}=97.5$ MeV,
$f_{\eta^{\prime}}=74.4$ MeV. In Figs. \ref{fig:pi}, \ref{fig:eta},
\ref{fig:eta1} and \ref{fig:etaC} the parametrizations (\ref{Fcleo}) with
parameters $\Lambda_{\pi}=776$ MeV, $\Lambda_{\eta}=800$ MeV, $\Lambda
_{\eta^{\prime}}=859$ MeV and $\Lambda_{\eta_{C}}=2.92$ GeV  are shown by
dot-dashed lines.

We see that the parametrization
(\ref{Fcleo}) describes the $\eta^{\prime}$ and $\eta_{C}$ mesons form factors
(Figs. \ref{fig:eta1}, \ref{fig:etaC}) well at all measured momentum transfer.
However, it starts to deviate from data
for the pion and probably for the eta mesons form factors at momentum transfers squared larger than 10 GeV$^2$ (Figs.
\ref{fig:pi}, \ref{fig:eta}). Namely, the high momentum transfer
data obtained by BABAR collaboration show evident growth of the form factor
multiplied by $Q^{2}$ at large $Q^{2}$ for the pion and probably for the eta.
This is in contradiction with asymptotic formula (\ref{Fas}) and unexpected from the QCD
factorization approach \cite{Mikhailov:2009kf}\footnote{Small radiative
corrections can be attributed to slight changes in the parameters $\Lambda
_{P}$ in (\ref{Fcleo}).}. The $\eta_{c}$ form factor is in good agreement with
 predictions \cite{Feldmann:1997te}.

In \cite{Dorokhov:2010bz,Dorokhov:2010zzb} the asymptotic of the pseudoscalar
meson transition form factor generalizing the result (\ref{Fas}) was derived
\begin{equation}
F_{P\gamma^{\ast}\gamma}\left(  0;Q^{2}\rightarrow\infty,0\right)
=\frac{2f_{P}}{3}\int_{0}^{1}dx\frac{\varphi_{P}^{f}\left(  x\right)
}{D\left(  xQ^{2}\right)  },\label{FpqcdNL}%
\end{equation}
where $D\left(  k^{2}\right)  $ is the nonperturbative inverse quark
propagator with property $D\left(  k^{2}\rightarrow\infty\right)  \rightarrow
k^{2}$ and $\varphi_{\pi}\left(  x\right)  $ is the
meson distribution amplitude (DA) given in the nonlocal model by (see for
notations and the $\alpha$ representation \cite{Dorokhov:2010bz,Dorokhov:2010zzb})
\begin{equation}
\varphi_{P}\left(  x\right)  =\frac{N_{c}}{4\pi^{2}f_{\mathrm{PS},\pi}^{2}%
}\int_{0}^{\infty}\frac{dL}{L}e^{x\overline{x}LM_{P}^{2}}\left(
xG_{m,0}\left(  xL,\overline{x}L\right)  +\overline{x}G_{0,m}\left(
xL,\overline{x}L\right)  \right)  ,\label{PionDAsym}%
\end{equation}
with $\overline{x}=\left(  1-x\right)  $ and
\[
\int_{0}^{1}dx\varphi_{P}\left(  x\right)  =1.
\]
The properties of the form factor at large $Q^{2}$ strongly depend on the end
point behavior of the meson DA. In
\cite{Dorokhov:2009dg,Radyushkin:2009zg,Polyakov:2009je} it was assumed that
the pion is almost pointlike and that its DA can be almost constant (flat) in its shape. In this case the integrand in (\ref{FpqcdNL}) is poorly
convergent and the $1/Q^{2}$ behavior of the form factor is enhanced by
logarithmic factor $\ln(Q^{2}/M_{q}^{2})$ asymptotically or in rather wide region of large $Q^2$, where $M_{q}$ is dynamical
(constituent) quark mass. At the same time the DA with suppressed end point
behavior lead to the standard result (\ref{Fcleo}).

The solid line in Fig. \ref{fig:pi} is the pion transition form factor
calculated in the nonlocal chiral quark model \cite{Dorokhov:2010bz,Dorokhov:2010zzb} with
the mass parameters $M_{Q}=135$ MeV and strange quark mass $M_{s}=250$ well describes the BABAR data. There are
other attempts to explain the BABAR data on the pion form factor within
different factorization schemes
\cite{Li:2009pr,Kochelev:2009nz,Bystritskiy:2009bk,Kroll:2010bf,Wu:2010zc,Brodsky:2011xx,Brodsky:2011sk}.

By using singlet-octet mixing scheme
one has for $\eta$ and $\eta^{\prime}$ meson form factors
\cite{Dorokhov:2011zf}
\begin{align}
& \mathrm{F}_{\eta\gamma^{\ast}\gamma}\left(  m_{\eta}^{2};Q^{2},0\right)
=\frac{1}{3\sqrt{3}}\left[  \left(  5F_{u}\left(  m_{\eta}^{2};Q^{2},0\right)
-2F_{s}\left(  m_{\eta}^{2};Q^{2},0\right)  \right)  \cos\theta(m_{\eta}%
^{2})-\right.  \nonumber\\
& \left.  -\sqrt{2}\left(  5F_{u}\left(  m_{\eta}^{2};Q^{2},0\right)
+F_{s}\left(  m_{\eta}^{2};Q^{2},0\right)  \right)  \sin\theta(m_{\eta}%
^{2})\right]  ,\nonumber
\end{align}%
\begin{align}
& \mathrm{F}_{\eta^{\prime}\gamma^{\ast}\gamma}\left(  m_{\eta^{\prime}}%
^{2};Q^{2},0\right)  =\frac{1}{3\sqrt{3}}\left[  \left(  5F_{u}\left(
m_{\eta^{\prime}}^{2};Q^{2},0\right)  -2F_{s}\left(  m_{\eta^{\prime}}%
^{2};Q^{2},0\right)  \right)  \sin\theta(m_{\eta^{\prime}}^{2})+\right.
\nonumber\\
& \left.  +\sqrt{2}\left(  5F_{u}\left(  m_{\eta^{\prime}}^{2};Q^{2},0\right)
+F_{s}\left(  m_{\eta^{\prime}}^{2};Q^{2},0\right)  \right)  \cos
\theta(m_{\eta^{\prime}}^{2})\right]  ,
\end{align}
From these expressions it is clear that in general it is not possible to
separate the nonstrange and strange components of the form factors as it was
attempted to do in \cite{:2011hk} and then used in some theoretical works.

The solid line in Figs. \ref{fig:eta}, \ref{fig:eta1}, is the eta and eta prime transition form factors
calculated in the nonlocal chiral quark model \cite{Dorokhov:2010bz,Dorokhov:2010zzb} with
the mass parameters $M_{Q}=135$ MeV and strange quark mass $M_{s}=250$ well describes the BABAR data.

\begin{figure}[th]
\hspace*{-10mm}\includegraphics[width=0.6\textwidth]{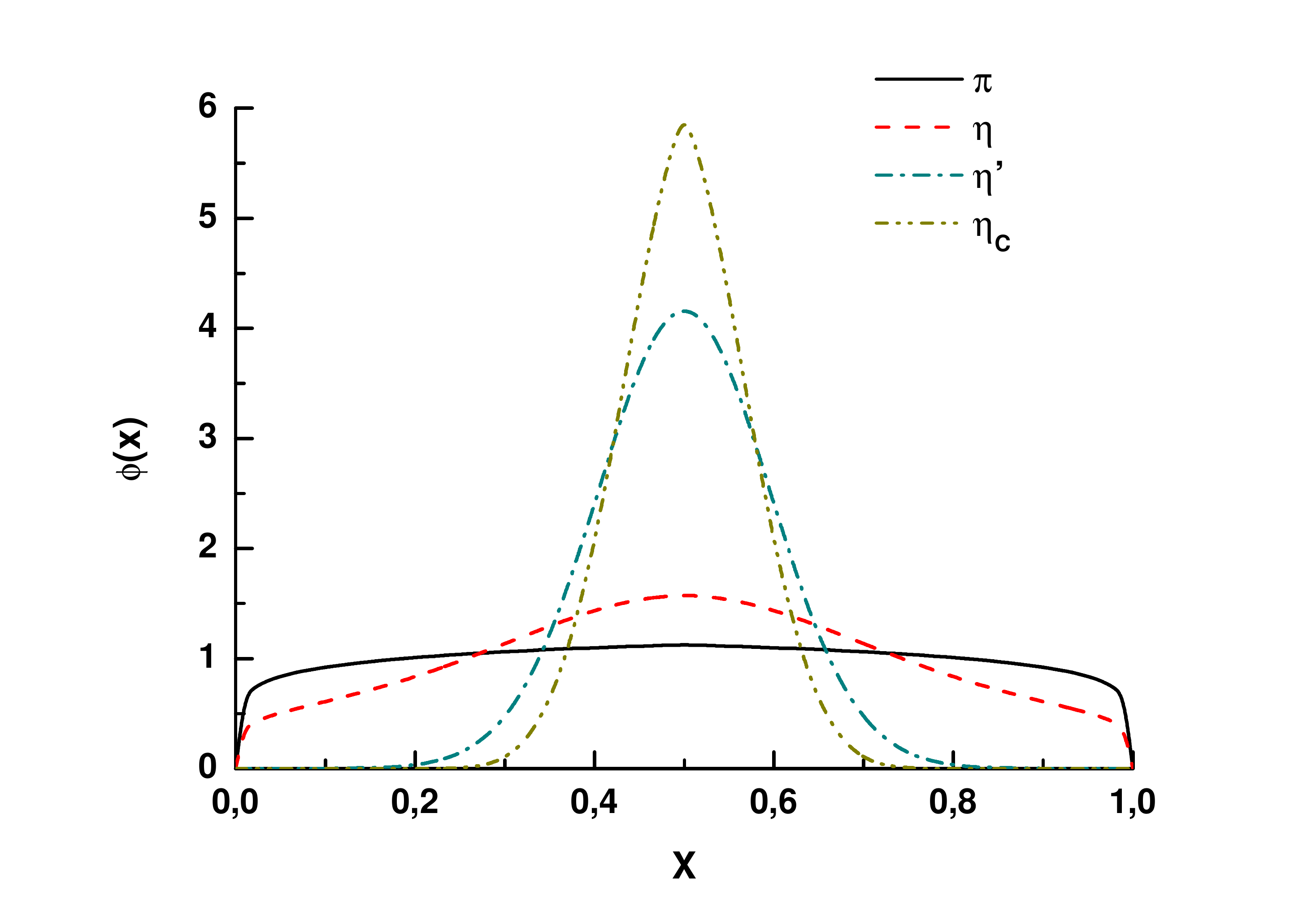} \vspace
*{-10mm}\caption{{\protect\footnotesize The pion, }$\eta,\eta^{\prime}%
,${\protect\footnotesize  and }$\eta_{C}${\protect\footnotesize  distribution
amplitudes for the nonlocal chiral quark model with parameters }$M_{u}=135$%
{\protect\footnotesize  MeV, }$M_{s}=250$ MeV, $M_{c}=1550${\protect\footnotesize
MeV, }}%
\label{PhiXmodel}%
\end{figure}

From above results one may conclude that the possible origin of the
difference of the asymptotic behavior of the pion and $\eta$ from one side and of the $\eta^{\prime}$
meson form factors is the fact that
$\eta^{\prime}$ is much heavier than the pion and $\eta$. We have checked that it is this
property responsible for the change of the almost flat DA  in the pion and $\eta$ cases  for the suppressed at end points DA in the cases of the $\eta^{\prime},\eta_c$ mesons (Fig.
\ref{PhiXmodel}) \cite{Dorokhov11}.

\section{Pion transversity form factors}

The \emph{transversity} form factors (TFFs) of the pion provide valuable
insight into chirally-odd generalized parton distribution functions (GPDs) as
well as into the non-trivial spin structure of the pion. These interesting
quantities have been determined for the first time on the
lattice~\cite{Brommel:2007xd}. Formally, the TFFs, denoted as $B_{Tni}^{\pi
}(t)$, are defined as
\begin{equation}
\langle\pi^{+}(P^{\prime})|\mathcal{O}_{T}^{\mu\nu\mu_{1}\cdots\mu_{n-1}}%
|\pi^{+}(P)\rangle=\mathcal{AS}\,{\bar{P}^{\mu}\Delta^{\nu}}\sum_{{i=0,{even}%
}}^{n-1}\Delta^{\mu_{1}}\cdots\Delta^{\mu_{i}}\bar{P}^{\mu_{i+1}}\cdots\bar
{P}^{\mu_{n-1}}\frac{B_{Tni}^{\pi_{,}u}(t)}{m_{\pi}},\label{def}%
\end{equation}
where $P^{\prime}$ and $P$ are the momenta of the pion, the $\bar{P}=\frac
{1}{2}(P^{\prime}+P)$, $\Delta=P^{\prime}-P$, and $t=\Delta^{2}$. The symbol
$\mathcal{AS}$ denotes symmetrization in $\nu,\ldots,\mu_{n-1}$, followed by
antisymmetrization in $\mu,\nu$, with the additional prescription that traces
in all index pairs are subtracted. The tensor operators are given by
\[
\label{eq:gff:def}\mathcal{O}_{T}^{\mu\nu\mu_{1}\cdots\mu_{n-1}}%
=\mathcal{AS}\;\overline{u}(0)\,i\sigma^{\mu\nu}i{D}^{\mu_{1}}\dots i{D}%
^{\mu_{n-1}}u(0),
\]
where $D=\frac{1}{2}\left(  \overrightarrow{D}-\overleftarrow{D}\right)  $
denoting the QCD covariant derivative.

The available full-QCD lattice results \cite{Brommel:2007xd} are for
$B^{\pi,u}_{10}$ and $B^{\pi,u}_{20}$ and for $-t$ reaching 2.5~GeV$^{2}$,
with moderately low values of the pion mass $m_{\pi}\sim600$~MeV. The
calculation uses the set of QCDSF/UKQCD $N_{f} = 2$ improved Wilson fermion
and the Wilson gauge-action ensembles.

\begin{figure}[th]
\begin{minipage}[c]{8cm}
\includegraphics[width=1.0\textwidth]{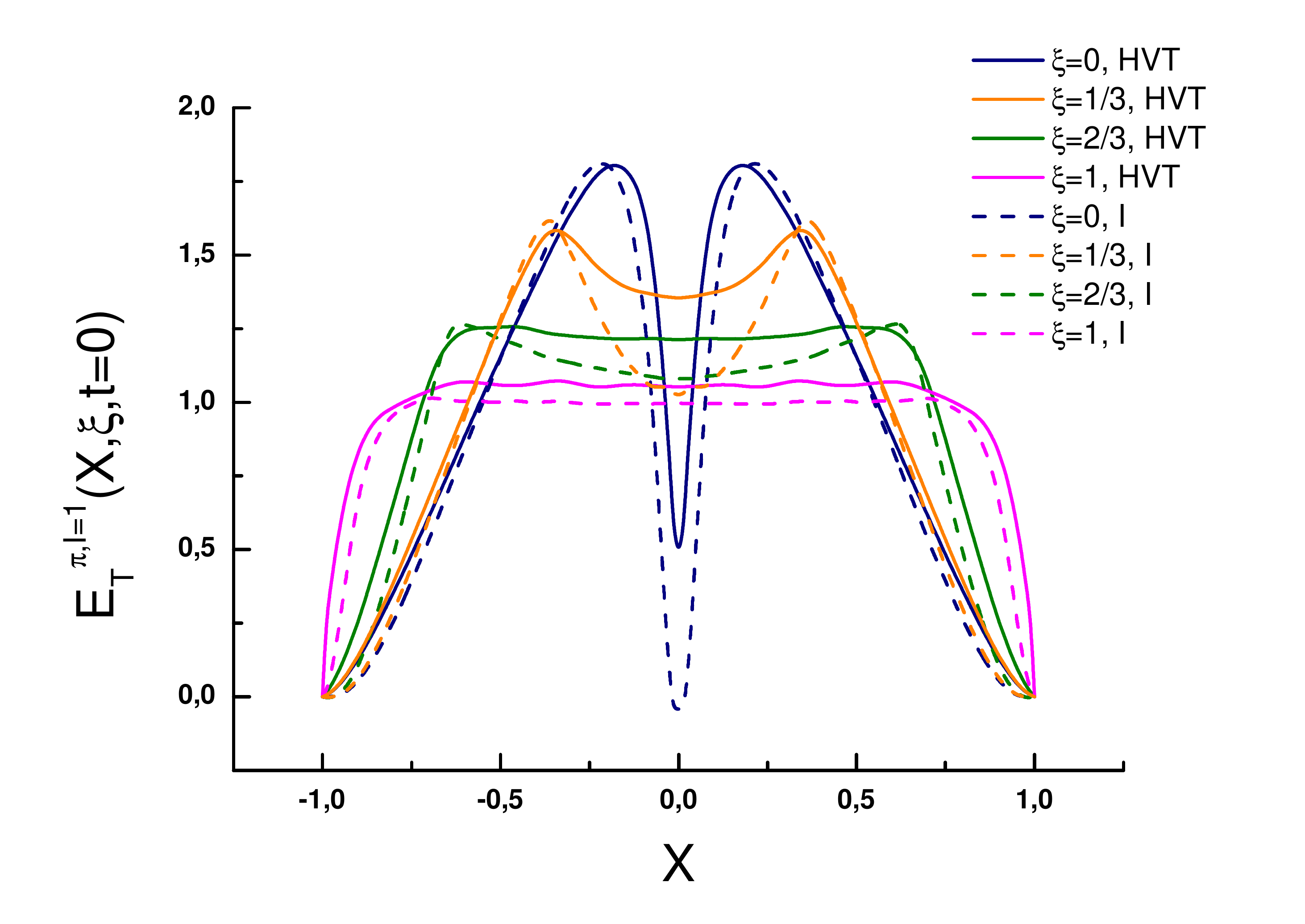}
\end{minipage}
\begin{minipage}[c]{8cm}
\includegraphics[width=1.0\textwidth]{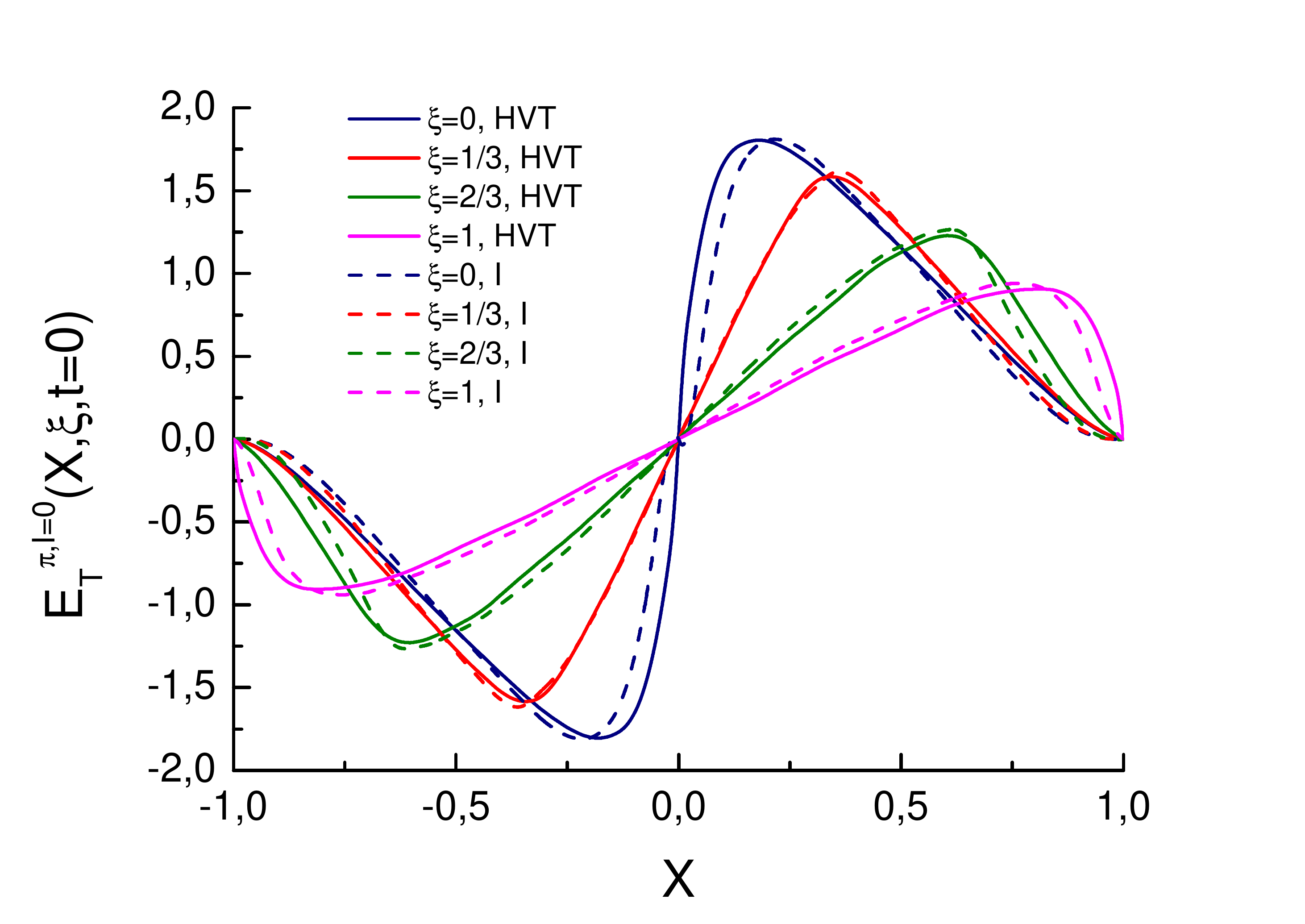}
\end{minipage}
\caption{{\protect\footnotesize The pion tGPD for isovector and isoscalar
cases in the HTV model (solid lines) and in the instanton model (dashed lines)
for several values of $\xi$.}}%
\label{E(x)I}%
\end{figure}

Within the effective nonlocal model one gets from the triangle diagram
corresponding to (\ref{def}) the transversity pion form factors (see
\cite{Broniowski:2010nt,Dorokhov:2011ew}, where notations are explained)
\begin{align}
& B_{Tni}^{u}\left(  t\right)  =\frac{N_{c}}{4\pi^{2}f_{\pi}^{2}}\frac{\left(
n-1\right)  !}{i!\left(  n-1-i\right)  !}\int\frac{d\left(  \alpha\beta
\gamma\right)  }{\Delta^{n+2}}\beta^{n-1-i}\left(  \frac{\gamma-\alpha}%
{2}\right)  ^{i}e^{-\frac{\alpha\gamma}{\Delta}t}\nonumber\\
& \left[  2\alpha G_{m,0,0}\left(  \alpha,\beta,\gamma\right)  +\beta
G_{0,m,0}\left(  \alpha,\beta,\gamma\right)  \right]  \!\!,
\end{align}
where $i=0,2,...\leq n-1$ and pion transversity generalized parton
distribution (Fig. \ref{E(x)I})
\begin{align}
& E_{T}^{\pi}\left(  X,\xi,t\right)  =\Theta\left(  X+\xi\right)  \frac{N_{c}%
}{4\pi^{2}f_{\pi}^{2}}\int_{0}^{\infty}\!\!\!\!\!d\gamma\int_{\mathrm{max}%
\left\{  0,\gamma\frac{\xi-X}{\xi+X}\right\}  }^{\infty}%
\!\!\!\!\!\!\!\!\!\!\!\!\!\!\!\!\!\!\!d\alpha\,e^{-\frac{\alpha\gamma}{\Delta
}t}\nonumber\\
& \frac{\alpha G_{m,0,0}\left(  \alpha,\beta,\gamma\right)  +\beta
G_{0,m,0}\left(  \alpha,\beta,\gamma\right)  +\gamma G_{0,0,m}\left(
\alpha,\beta,\gamma\right)  }{\Delta^{2}\left(  1-X\right)  },
\end{align}
where $\xi=-\frac{\left(  aq\right)  }{2\left(  aP\right)  }$, ($a^{2}=0$),
$\beta=\frac{\left(  X+\xi\right)  \alpha+\left(  X-\xi\right)  \gamma}{1-X}$,
$\Delta=\left[  \alpha+\gamma+\xi\left(  \alpha-\gamma\right)  \right]
/({1-X})$. These studies show how the spinless pion acquires a non-trivial
spin structure within the framework of chiral quark models.

\section{Conclusions}

The main conclusion is that the study of pion and other pseudoscalar mesons provides reach
information about dynamics of strong interactions and in some rare processes
testifies the standard model. New high statistics experiments in wider
kinematical region are urgent for further progress. It is expected new
interesting theoretical results of the nonperturbative QCD dynamics from the
Lattice QCD and effective chiral quark models.

We are grateful to the Organizers and personally S. Dubnicka, A.-Z.
Dubnickova, E. Bartos, M. Hnatic, N.Shumeiko, N.Maksimenko, V.Mossolov, S. Eidelman for nice meetings and kind invitation to
present our results.
This work is supported in part by RBFR grants 10-02-00368
and 11-02-00112.

\nocite{*}



\end{document}